\documentclass[showpacs,preprintnumbers,twocolumn,amsmath,amssymb]{revtex4}
\usepackage{graphicx}
\usepackage{dcolumn}
\usepackage{bm}

\begin{document}

\title{Andreev reflection and tunneling spectrum on metal-superconductor-metal junctions}
\author{W. LiMing}\email{wliming@scnu.edu.cn}
\author{J. J. Ouyang, Liangbin Hu}
\affiliation{Dept. of Physics, and Laboratory of Quantum Information Technology, School
of Physics and Telecommunication Engineering, South China Normal University, Guangzhou
510006, China} 
\begin{abstract}
The tunneling spectrum of an electron and a hole in metal-superconductor-metal junctions is computed using the Blonder-Tinkham-Klapwijk theory. The incident and the outgoing currents finally balance each other by an interface charge inside the superconductor and metal junction. The present computation shows a more abundant structure compared to that on a metal-superconductor junction, such as the resonance at bias voltages above the energy gap of the superconductor.  The density of the interface charge shows a quantum-like oscillation.
\end{abstract}
\date{\today}
\maketitle 
\section{Introduction}
The Andreev reflection on metal-superconductor junctions has a long history and still attracts great interests of scientists recently\cite{Blonder,Satoshi, Yang}. The tunneling spectrum of electrons on a metal-superconductor junction is sensitive to the energy gap of the superconductor, thus becoming an important technique to measure the gaps and gap properties of superconductors. When an electron transmits into a superconductor to form a Cooper pair there appears a hole reflection, which is called the Andreev reflection. A commonly used description to the Andreev reflection is the Blonder-Tinkham-Klapwijk(BTK) theory\cite{Blonder}, which take the interface in a metal -superconductor junction as a $\delta(x)$ potential barrier. Wei, Dong and Xing studied the tunneling spectrum in metal-superconductor-metal junctions(MSMJs) \cite{Wei} using the BTK theory. They found unreasonable results that the incident and outgoing currents in both sides do not balance each other. Thus they claimed that the BTK theory are not suitable for the Andreev reflection in MSMJs and treated them in the Landauer-Buttiker formalism\cite{Buttiker}.

The authors of this paper think that since the BTK theory uses the standard quantum mechanics and did not do any unreasonable approximation it should be still working well in MSMJs. The unbalanced currents must cause a charge accumulation on the interface in the MSMJs. These interface charges change the potential on the metals and finally reaches a balance. This dynamic process can be simulated by solving the time-dependent Sch\"{r}odinger equation. In the present paper a simpler model is set up to describe the final balanced state of the MSMJs. The differential conductance of MSMJs is computed in this model.

\section{formalism}
The metal-superconductor-metal junctions are shown in Fig.1, where two thin insulating layers exist inside the two junctions becoming two potential barriers.  An electron and a hole are incident into the junctions under a bias voltage from the left side and the right side, respectively. Due to the two barriers the electron and the hole are partially reflected and partially transmitted into the superconductor(SC). In the SC a pseudo-particle can only propagate above the energy gap $\Delta$. When the bias voltage is smaller than $\Delta$ only Cooper pairs can propagate in the superconductor. In this case there will be a hole reflected on both sides, respectively, which are called Andreev reflection in literatures.

\begin{figure}
\includegraphics[width=8cm,height=5cm]{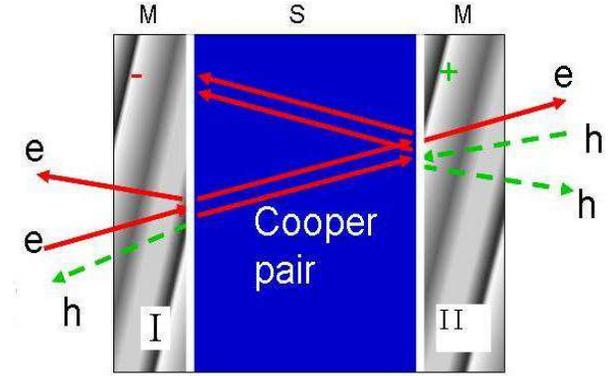}
\caption{A metal-superconductor-metal junction with an incident electron and an incident hole from the left side and the right side respectively under a bias voltage. The metal leyer and the superconductor are labeled by M and S, respectively.}
\end{figure}
A key idea of this paper is that the metal-superconductor interface at the right side will be charged due to the supercurrent in the SC.
The tunneling process of an electron and a hole on the MSMJs is described by the Bogoliubov-de Gennes (BdG) equation given by\cite{BdG}
\begin{align}
i\hbar {\partial \over \partial t} \begin{pmatrix} u({\bf r},t)\\v({\bf r},t)\end{pmatrix} = \begin{pmatrix} H_0 &\Delta \\ \Delta^* & -H_0\end{pmatrix} \begin{pmatrix} u({\bf r},t)\\v({\bf r},t)\end{pmatrix}\\
H_0 = -{\hbar^2\over 2m} \nabla^2 + V({\bf r}) -\mu
\end{align}
where $V({\bf r})= eV, -eV'$ in the left and right metals, respectively, and $0$ in the SC. On the interfaces $V({\bf r})= U\delta(x)$. $eV'=eV+e\delta V$ is the potential energy at the right side which is slightly changed by the interface charge. The continuity equation of the BdG equation is given by
\begin{align}
{\partial\rho_Q({\bf r})\over \partial t} + \nabla \cdot ({\bf J}_Q({\bf r}) +{\bf J}_S({\bf r})) = 0\end{align}
where $\rho_Q({\bf r}$ and ${\bf J}_Q({\bf r})={e\hbar \over m} Im(u^*\nabla u +  v^*\nabla v)$ are the charge density and the  current density of quasi-particles, respectively.  The supercurrent density is defined by $\nabla \cdot {\bf J}_S({\bf r}) = -{4e\over \hbar} Im(\Delta u^*({\bf r})v({\bf r}))
$. It can be proved that quasi-particle current is continuous along the whole MSMJs, but the supercurrent is obviously discontinuous since no supercurrent exists in a metal. This charges the interface like the charging process on a capacitor with a bias voltage. This can be seen by integrating the continuity equation over a slap volume across the interface, resulting in
\begin{align}
{\partial \sigma(t)\over \partial t} = J_S({x=L})
\end{align}
where $\sigma(t)$ is the surface charge density on the interface and $J_S(x=L)$ is the supercurrent at the interface between the SC and metal II. The interface charge density reaches saturation when the current densities on both sides of the interface balance. The interface charge distribution changes the potential energies of electrons and holes in  metal II thus the outgoing current is changed and finally reaches balance with the incident current.
\begin{figure}
\includegraphics[width=8cm,height=4cm]{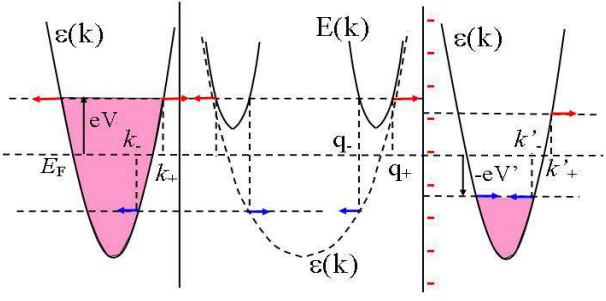}
\caption{The energy dispersion on the MSMJs in the case of $eV>\Delta$. The arrows label the propagating directions of quasi-electrons (red) and quasi-holes (blue). The charges on the interface are labeled by a column of "-", which can also be positive.}
\end{figure}

As seen from Fig.2 when the bias voltage satisfying $eV > \Delta$ an incident electron in metal I tunnels into the SC and propagate as a quasi-particle. It is then partially reflected by the following interface and partially tunnels into the following metal. So does an incident hole from the right side. The wave functions of the electron and hole are written as
\begin{align}
\psi_{I}= \left(
  \begin{array}{c}  e^{ik_{+}x}+be^{-ik_{+}x}\\   ae^{i k_{-}x}\\
  \end{array}\right) \label{wf1}
\\
\psi_{SC}=\biggl[ce^{iq_{+}x}
\left(
  \begin{array}{c}   u_{+}\\  v_{+}\\  \end{array} \right)  +de^{-iq_{-}x}
\left(
  \begin{array}{c}     u_{-}\\     v_{-}\\   \end{array}
\right)  \nonumber\\ + e e^{-iq_{+}x}
\left(   \begin{array}{c}     u_{+}\\    v_{+}\\   \end{array} \right) +fe^{iq_{-}x}
\left(   \begin{array}{c}  u_{-}\\   v_{-}\\   \end{array} \right)\biggr]
\\
\psi_{II}=\left( \begin{array}{c}  g e^{i k'_{+}x}\\
    e^{i k'_- x} + h e^{-i k'_- x}\\  \end{array}
\right)\label{wf3}
\end{align}
where $\left(
  \begin{array}{c}   u_{+}\\  v_{+}\\  \end{array} \right)$ and $\left(
  \begin{array}{c}   u_{-}\\  v_{-}\\  \end{array} \right)$ are the quasi-eletron and quasi-hole wave functions, respectively. They are given by
\begin{align}
u_{\pm} = \sqrt{{1\over 2} \pm {\epsilon_{q\pm}-\mu \over 2E}}, v_{\pm} = {|\Delta|\over \Delta}\sqrt{{1\over 2} \mp {\epsilon_{q\pm}-\mu\over 2E}}\\
E = \sqrt{(\epsilon_{q\pm}-\mu)^2 + |\Delta|^2} ,\quad \epsilon_{q\pm}={\hbar^2 q_\pm^2\over {2m}}
\end{align}
The wave vectors are determined by
 \begin{align}
E= eV={\hbar^2 k_+^2\over {2m}} - \mu =  \mu - {\hbar^2 k_-^2 \over {2m}}\\
\quad k_\pm/k_F = \sqrt{1\pm {E/\mu}}\\
\quad q_\pm/k_F = \sqrt{1\pm \sqrt{E^2-\Delta_q^2}/\mu}\\
-E' = -eV' ={\hbar^2 {k'}_-^2\over {2m}} - \mu =  \mu - {\hbar^2 {k'}_+^2 \over {2m}}\\
k'_\pm/k_F = \sqrt{1\pm {E'/\mu}}
\end{align}

In the case of $eV<\Delta$, $ q_\pm$ become complex numbers, so that the wave functions of electrons become damping traveling waves and Cooper pairs appear.

The constants $a,b,c,d,e,f,g,h$ in (\ref{wf1}) to (\ref{wf3}) are obtained from the boundary conditions of wave functions on the interfaces, where the wave functions are continuous but
the first derivatives of them have jumps due to the $\delta(x)$ barriers, i.e.,
\begin{align}
\psi_{I}(0^{-})=\psi_{SC}(0^{+})\\
\psi'_{I}(0^{-})-\psi'_{SC}(0^{+})+Zk_F\psi_{I}(0)=0\\
\psi_{SC}(L)=\psi_{II}(L)\\
\psi'_{SC}(L)-\psi'_{II}(L)+Zk_F\psi_{II}(L)=0
\end{align}
where $Z=\frac{2mU}{\hbar^{2}k_F}$ and $L$ is the thickness of the SC.
Finally the current density, the super-current and the differential conductance are computed with these constants, which are given by
\begin{align}
J_Q(I) = {2e}[k_+(1-|b(eV)|^2)+k_-|a(eV)|^2]\\
J_Q(II) = {2e}[k'_-(1-|h(eV)|^2)+k'_+|g(eV)|^2]
\\
{dI_1(eV)\over dV}\sim 1-|b(eV)|^2+|a(eV)|^2\\
{dI_2(eV)\over dV}\sim 1-|h(eV')|^2+|g(eV')|^2
\end{align}

\section{Results of computation}
To view the effect of the interface charges we first compute the current densities in the MSMJs with and without interface charges. As shown in Fig.3(left), the charge currents is continuous but the incident and outgoing current do not balance with each other.  It is natural to think that the interface at the right side will be charged until saturation so that both the currents balance. This is true. As seen in Fig.3(right)  the two currents equal to each other after fully charged. It is interesting to note that the super-currents are always complementary to the charge currents in both cases so that the total currents are always constant in the SC. In addition, the quasi-particle current and the super-current with interface charges become symmetric about the center of the SC. This result convinces us that the interface charges have a true effect on the currents in the MSMJs.

\begin{figure}
\includegraphics[width=4cm,height=3.5cm]{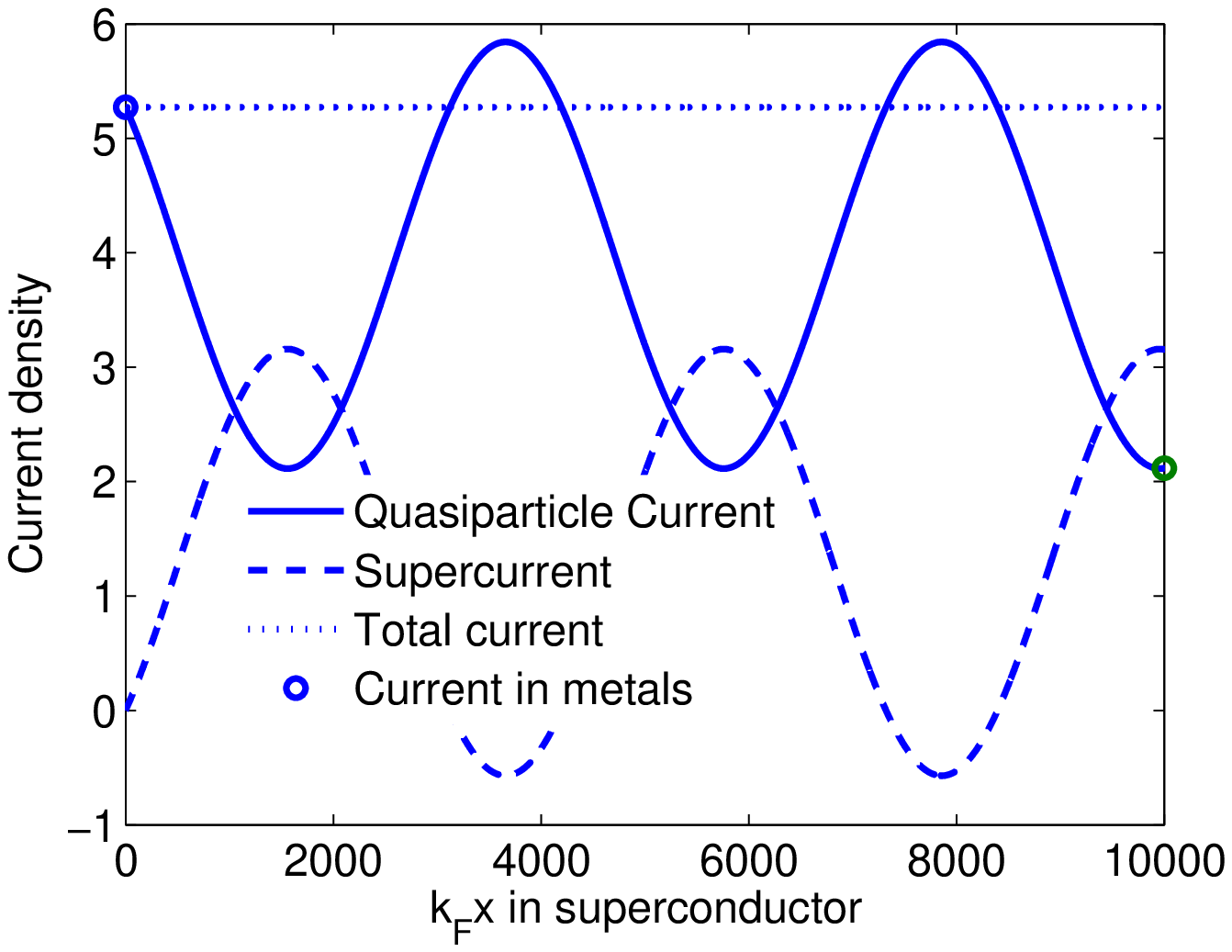}
\includegraphics[width=4cm,height=3.5cm]{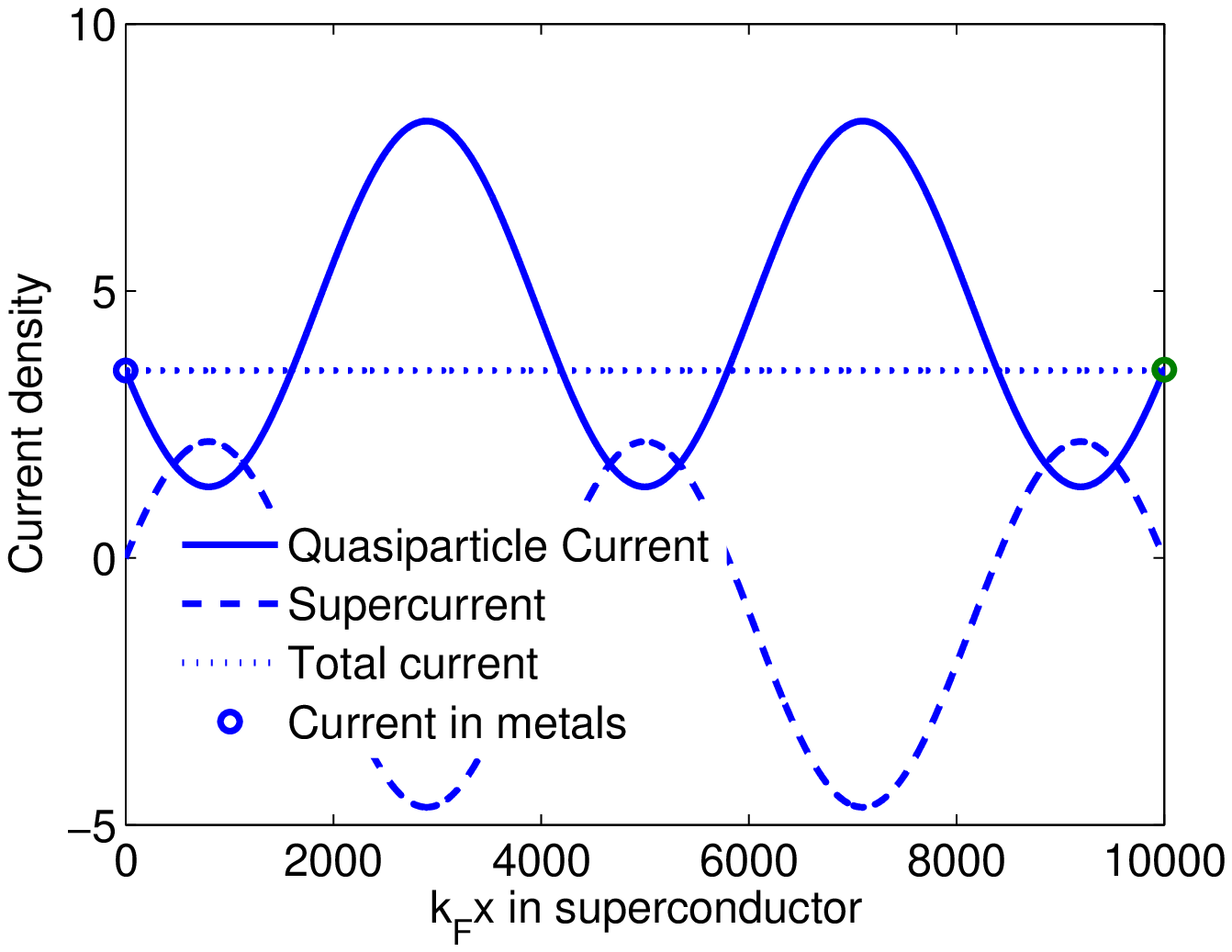}
\caption{Charge current densities inside the SC without(left, $\sigma= 0.$) and with(right,$\sigma= 0.24\times 10^{-4}e$) interface charging. The parameters are set to $\mu=0.5, \Delta = 0.001\mu, Z = 1, k_F L=10000, eV = 1.8\Delta$. Circles label the incident and outgoing currents in the metals. }
\end{figure}

Further computation shows that the  the incident and outgoing currents in fact oscillate on the interface charges.  There exists a series of interface charge densities  which balance the incident and outgoing currents, as shown by the matching points in Fig.4. The physics of this periodicity is still an open question. One may find a quantum behavior of the interface charge densities.
\begin{figure}
\includegraphics[width=5cm,height=3.5cm]{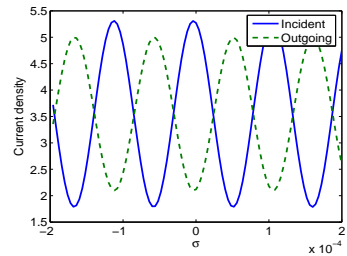}
\caption{The incident and outgoing current densities versus different interface charge densities at $eV =1.8\Delta$. Other parameters see Fig.3. }
\end{figure}

The tunneling spectrum, i.e. the normalized differential conductivities at different bias voltages without and with interface charges are shown in Fig.5. Without interface charges the incident and outgoing currents match each other only under small bias voltages. After fully charged the two currents overlap in the whole bias region. Therefore, the conservation of currents in the MSMJs is realized by means of the interface charging. In addition, the probability flux  $|a|^2+|b|^2+|f|^2+|g|^2=2$ is always conserved for different bias voltages.

\begin{figure}\includegraphics[width=4cm,height=3.5cm]{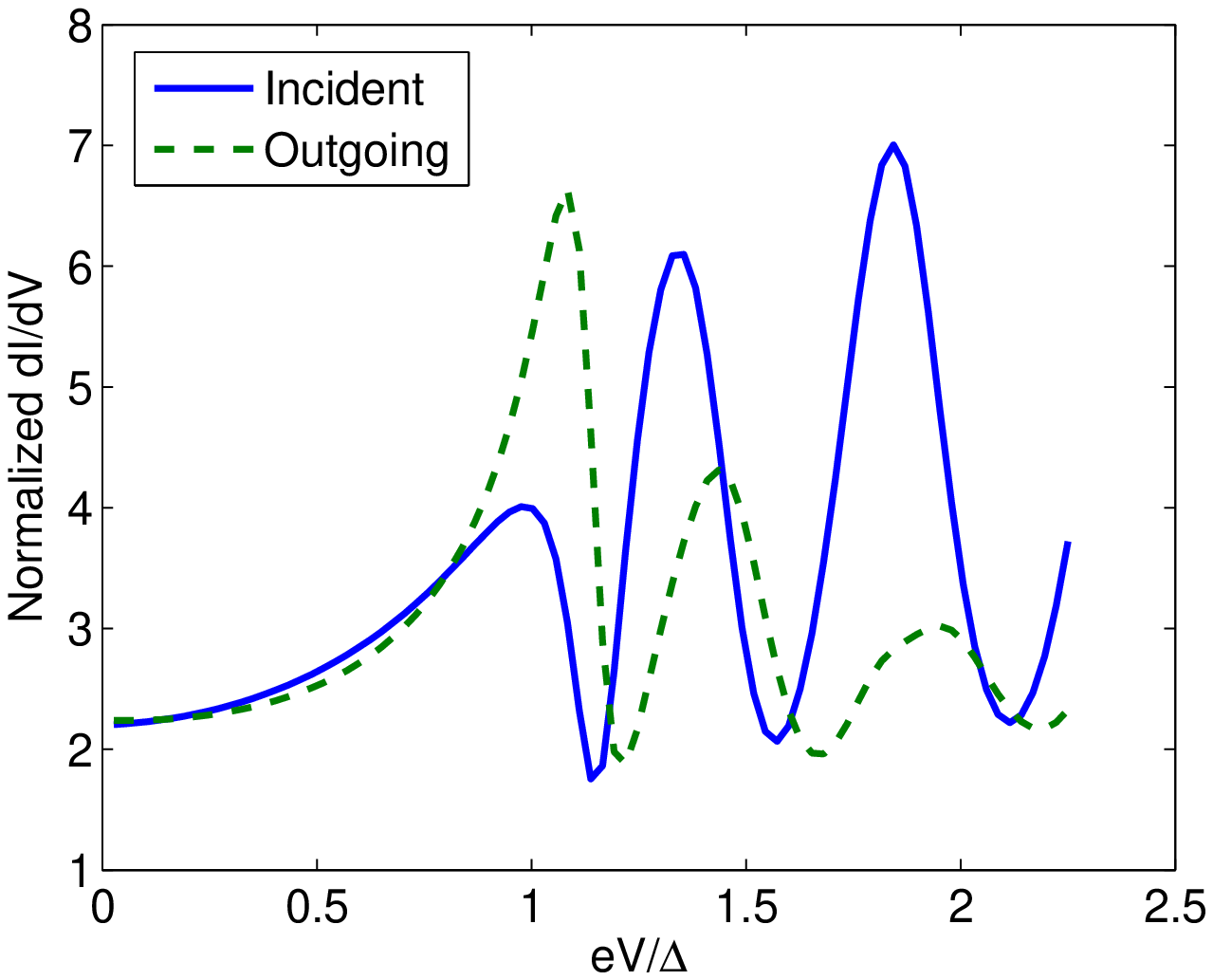}
\includegraphics[width=4cm,height=3.5cm]{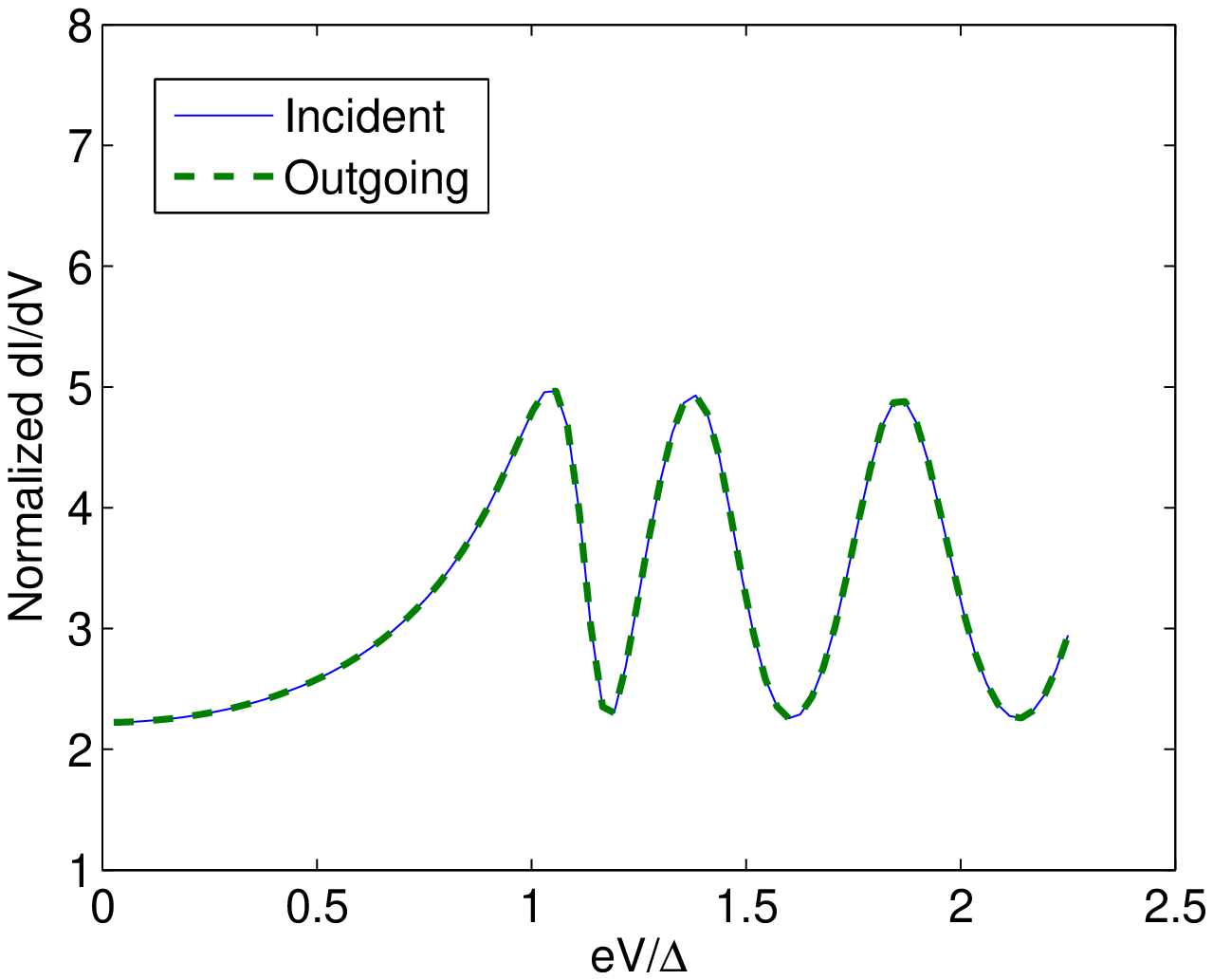}
\caption{The normalized differential conductance without(left panel) and with (right panel) interface charges versus bias potential. Parameters see Fig.3. }
\end{figure}
\begin{figure}
\includegraphics[width=4cm,height=3.5cm]{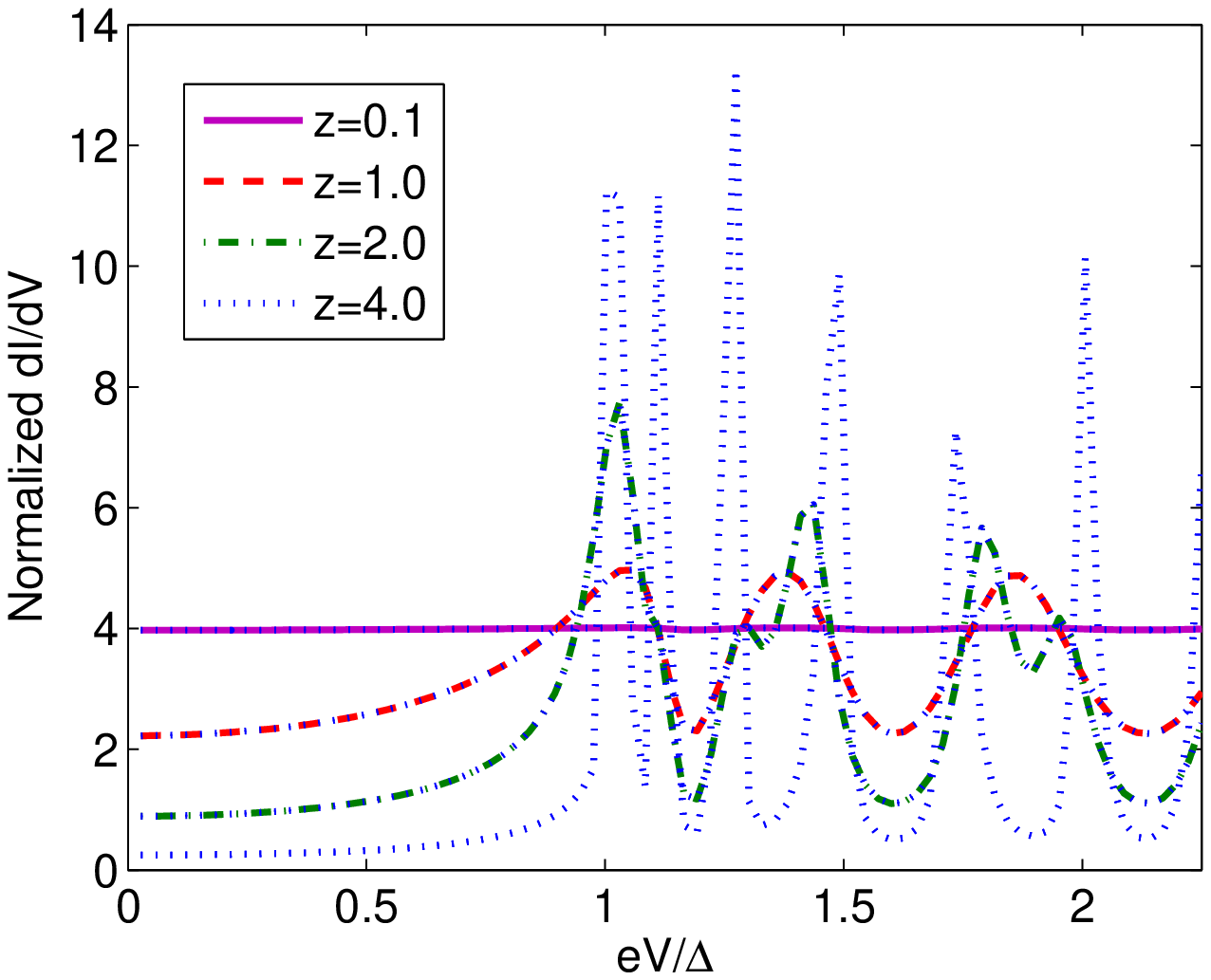}\includegraphics[width=4cm,height=3.5cm]{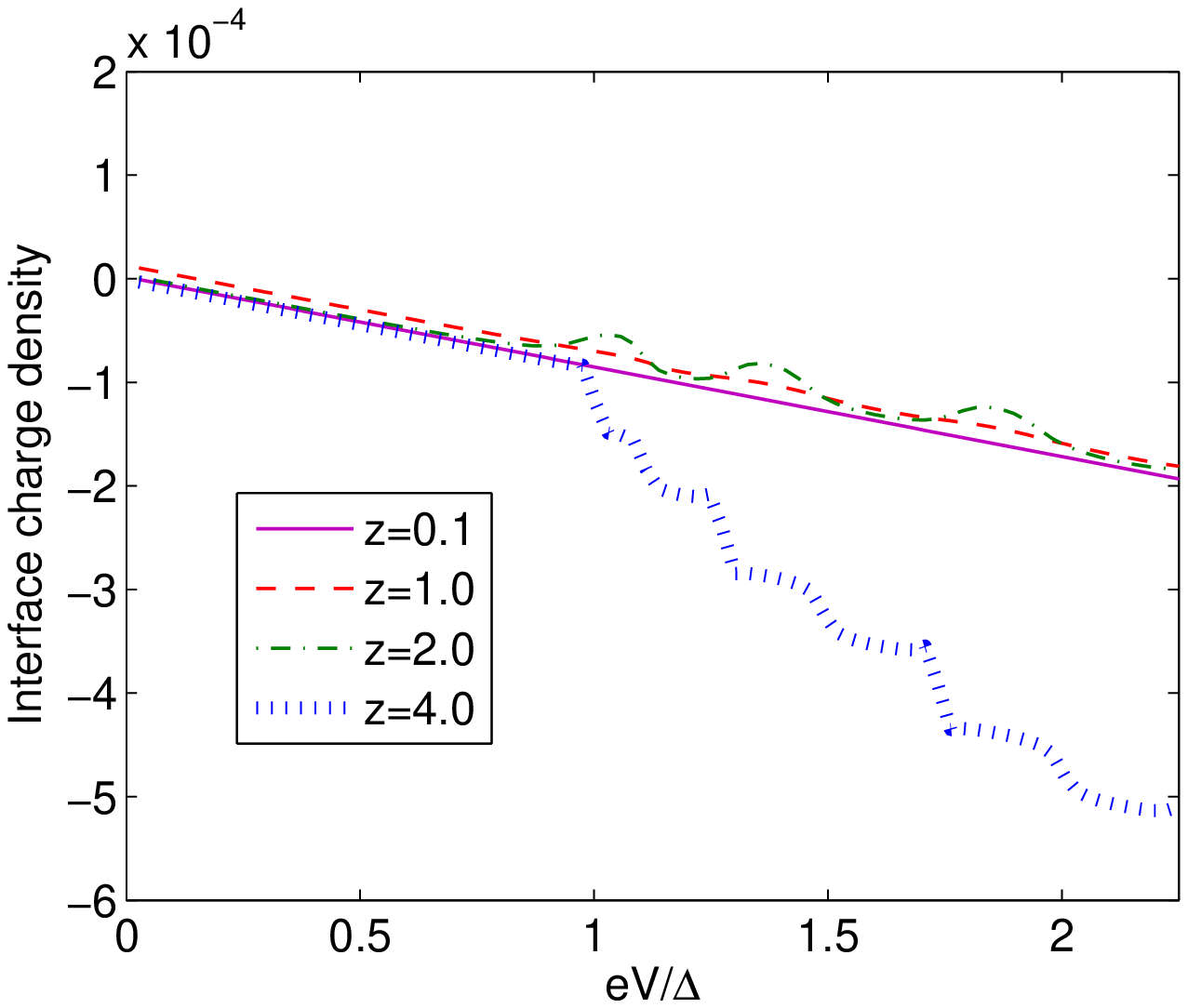}
\caption{The differential conductance (left) and the interface charge density (right) versus the bias voltage for different interface barriers. Other parameters see Fig.3. }
\end{figure}
Finally the tunneling spectra for the differential electric conductance (DEC) across different interface potential barriers are computed, as shown in Fig.6(left). Across a vanishing barrier the DEC approaches to a constant 4, which originates from the electron and hole injection. This is very different from that across a metal-superconductor junction, where the DEC drops gradually from a constant value to about half at the bias voltage of the energy gap($\Delta$). The main difference between these two cases is that there is a hole injection from the right side across the MSMJs. As the interface barrier increases the DEC shows  a few sharper and sharper resonances. Their positions depends on the width of the superconductor inside the MSMJs. Below the bias voltage $e\Delta$ the DEC decreases rapidly similar to the case across a metal-superconductor junction. This provides a measurement to the energy gap of the superconductor. The interface charge shows an interesting transition at the gap when the interface barrier reaches 4, see Fig.6(right). This may be taken as a signal that the resonance is fully set up.

In summary, the tunneling spectrum of an electron and a hole on MSMJs is computed through the Bogoliubov-de Gennes (BdG) equation. It shows a more abundant structure compared to that on a metal-superconductor junction, such as the resonance above the gap voltage. An important effect occurring in the present device is that the incident current and the outgoing current balance each other only when there is an interface charge inside the superconductor and metal junction.  The density of this interface charge shows a quantum-like oscillation, where a series of densities balance the currents.

\section{acknowledgment}
 This work was supported by the National Natural Science Foundation of
China (Grant No. 10874049),  the State Key Program for Basic Research of China (No.
2007CB925204) and the Natural Science Foundation of Guangdong province ( No. 07005834 ).

\end{document}